\documentclass[12pt]{iopart}
\usepackage{iopams}
\usepackage{setstack}
\usepackage[dvips]{graphicx}

\begin{document}

\title[Entanglement and squeezing in a two-mode system]{Entanglement and squeezing in a two-mode system: theory and experiment}

\author{V. Josse, A. Dantan, A. Bramati and E.
Giacobino}

\address{Laboratoire Kastler Brossel, Universit\'{e} Pierre et Marie Curie,
4 place Jussieu, F75252 Paris Cedex 05, France}

\begin{abstract}
We report on the generation of non separable beams produced via
the interaction of a linearly polarized beam with a cloud of cold
cesium atoms placed in an optical cavity. We convert the squeezing
of the two linear polarization modes into \textit{quadrature
entanglement} and show how to find out the best entanglement
generated in a two-mode system using the inseparability criterion
for continuous variable [Duan \textit{et al.}, Phys. Rev. Lett.
\textbf{84}, 2722 (2000)]. We verify this method experimentally
with a direct measurement of the inseparability using two homodyne
detections. We then map this entanglement into a polarization
basis and achieve \textit{polarization entanglement}.
\end{abstract}


\newcommand{\beq}{\begin{equation}}
\newcommand{\eeq}{\end{equation}}
\newcommand{\beqr}{\begin{eqnarray}}
\newcommand{\eeqr}{\end{eqnarray}}
\newcommand{\lb}[1]{\label{#1}}
\newcommand{\ct}[1]{\cite{#1}}
\newcommand{\dagg}{\dagger}
\newcommand{\iab}{\mathcal{I}_{a,b}}
\newcommand{\iabt}{\mathcal{I}_{a,b}(\theta)}
\newcommand{\iabs}{\mathcal{I}_{\alpha,\beta}^S}
\newcommand{\daa}{\delta A_a}
\newcommand{\dab}{\delta A_b}
\newcommand{\dax}{\delta A_x}
\newcommand{\dai}{\delta A_y}
\newcommand{\dau}{\delta A_u}
\newcommand{\dav}{\delta A_v}
\newcommand{\dacp}{\delta A_{+}}
\newcommand{\dacm}{\delta A_{-}}


\section{Introduction}

With the recent progress in the quantum information field, there
has been a lot of interest in entanglement in the continuous
variable (CV) regime. Criteria to demonstrate and quantify CV
entanglement have been developed \ct{reid,duan} and experimentally
tested \ct{furusawa,polzik,zhang,silberhorn,bowen02prl,josse3}. In
particular, the possibility to map a quantum polarization state of
light onto an atomic ensemble \ct{polzik2} has stirred a great
deal of attention to the quantum features of polarized bright
beams. The notion of polarization entanglement, i.e. entanglement
between Stokes parameters of two spatially separated beams, has
been investigated by Korolkova \textit{et al.} \ct{korolkova} and
first demonstrated by Bowen \textit{et al.} \ct{bowen02prl} by
mixing two independent squeezed beams produced by OPAs.
Polarization entanglement was also achieved via the Kerr non
linearity of optical fibers \ct{glockl} and cold atoms
\ct{josse3}. These experiments are important steps in connection
with quantum teleportation \ct{treps}, quantum dense coding
\ct{li}, entanglement swapping \ct{glocklpra} and, more generally,
characterizing
entanglement in the CV regime \ct{bowen03prl}.\\
In this paper, we report on the generation of non separable beams
via the interaction of a linearly polarized light beam with a
cloud of cold cesium atoms placed in an optical cavity
\ct{josse3}. In previous works \ct{josse1,josse2}, we have shown
that, after the non linear interaction with the atoms, two modes
of the light exiting the cavity were squeezed: the mean field
mode, but also the orthogonally polarized vacuum. We develop here
a general method to find out the best entanglement - as measured
with the inseparability criterion \ct{duan} - produced in a two
mode system and characterize the correlation properties of the
system in the Poincar\'{e} sphere. The main result is that the
maximal entanglement corresponds to the sum of the minimum noises
of two "uncorrelated" modes. The maximally entangled modes are
then circularly polarized with respect to these modes. We stress
the similarity with the usual entanglement experiments, which mix
independent squeezed beams on a beamsplitter
\ct{bowen02prl,glockl}. Moreover, we show that this mixing of two
independent beams is equivalent to rotating the polarization basis
of a single beam exhibiting correlations between polarization
modes. This intuitive approach could be of interest for the study
of systems in which quantum correlations exist between
polarization modes, and allows one to think in terms of
independent beams mixing.\\
We then apply these results to our experiments \ct{josse3}, show
experimental evidence of both quadrature entanglement and
polarization entanglement. In Sec. \ref{quadentang} quadrature
entanglement is demonstrated by figuring out the maximally
entangled modes and checking the inseparability criterion between
these modes in a direct detection scheme. In Sec. \ref{polaentang}
we map the entanglement into a polarization basis via the mixing
of our quadrature entangled modes with an intense coherent beam,
the phase of which is locked to that of the first beam. We
therefore measure directly the Stokes parameters fluctuations of
the two spatially separated beams, thus demonstrating polarization
entanglement.

\section{Looking for maximal entanglement}\label{optentang}

\subsection{General method}
In this section we develop a general method to find out the
maximal entanglement in a two-mode system. We start with a "black
box" - in our case the atomic medium in the cavity - out of which
comes a light beam with unknown quantum properties. Let us stress
that the goal of this Section is to develop a method to
characterize quantum properties, such as entanglement and
squeezing, which have been previously created between some
polarization modes by some interaction. Let us denote by $A_a$ and
$A_b$ two orthogonally polarized modes of this beam. They satisfy
the standard bosonic commutation relations
$[A_{\alpha},A_{\beta}^{\dagger}]=\delta_{\alpha\beta}$. The usual
quadrature operators, with angle $\theta$ in the Fresnel
representation,

\beqr \nonumber X_{\alpha}(\theta)=A_{\alpha}
e^{-i\theta}+A^{\dagger}_{\alpha}e^{i\theta},\hspace{0.3cm}
Y_{\alpha}(\theta)=X_{\alpha}(\theta+\pi/2)\hspace{0.5cm}(\alpha=a,b)
\eeqr

are the continuous variable analogous of the EPR-type operators as
introduced by Einstein, Podolsky and Rosen \ct{einstein}. The
criterion derived by Duan \textit{et al.} and Simon \ct{duan} sets
a limit for inseparability on the sum of these EPR-type operators
variances

\beq \mathcal{I}_{a,b}(\theta)=
\frac{1}{2}\left[\delta(X_a+X_b)^2(\theta)+
\delta(Y_a-Y_b)^2(\theta)\right]<2 \label{critere} \eeq

For Gaussian states, $\mathcal{I}_{a,b}(\theta)<2$ is a sufficient
condition for entanglement and has already been used several times
to quantify continuous variable entanglement
\ct{korolkova,treps,glockl,josse3}. In this Section we look for
the best entanglement produced in the system: using unitary
transformations we therefore seek to minimize $\mathcal{I}_{a,b}$
with respect to $a$, $b$ and $\theta$. Expanding (\ref{critere}),
one gets

\beq \iabt=
\langle\daa^{\dagg}\daa+\daa\daa^{\dagg}+\dab^{\dagg}\dab+\dab\dab^{\dagg}\rangle
+4|\langle \daa\dab\rangle|\cos[2(\theta-\theta_{a,b})] \eeq

where $\theta_{a,b}$ is the phase of $\langle\daa\dab\rangle$. The
minimum value is reached for $\theta=\theta_{a,b}\pm\pi/2$:

\beq \iab=
\min_{\theta}\iabt=\langle\daa^{\dagg}\daa+\daa\daa^{\dagg}+\dab^{\dagg}\dab+\dab\dab^{\dagg}\rangle\;-\;
4|\langle \daa\dab\rangle|\label{valeurduan}\eeq

$\iab$ does not depend on local unitary operations performed
separately on $a$, $b$. It thus provides a good measurement of the
entanglement between modes $a$ and $b$ and will be used throughout
this paper. Consequently, one has to look for the polarization
basis ($a^*,b^*$) of the "maximally entangled modes" which
minimizes $\iab$. It is easy to see that the first term in
(\ref{valeurduan}) is independent of the polarization basis, since
it is the trace of the correlation matrix of modes $a$ and $b$.
The entanglement between $a$ and $b$ is therefore completely
determined by the correlation term $|\langle \daa\dab\rangle|$.\\
In order to find the strongest correlations we turn to a
particular basis for the fluctuations (we are only interested here
in what happens to the "noise ellipsoid", regardless of the mean
field): as shown in the Appendix A, there always exists two
orthogonally polarized modes $A_u$ and $A_v$ such that $\langle
\dau \dav\rangle=0$. The $u$, $v$ modes are "uncorrelated" in the
sense of the inseparability criterion and satisfy
$\mathcal{I}_{u,v}=\max_{a,b} \iab\geq 2$. Note that these modes
are not uncorrelated \textit{stricto sensu}, since $\langle
\dau\dav^{\dagger}\rangle$ can be non zero. Moreover, our choice
is not unique, since any $A_u'=e^{i\theta_u}A_u$ and
$A_v'=e^{i\theta_v}A_v$ also satisfy the same property. To
unambiguously determine the "uncorrelated" basis we choose modes
$u$ and $v$ such that $\langle \delta A_u^2\rangle$ and $\langle
\delta A_v^2\rangle$ are positive numbers. Physically, it means
that we choose $u$ and $v$ such that
their noise is minimum for the same quadrature $Y$. \\
Two orthogonally polarized modes $a$ and $b$ decompose on such a
basis

\beqr A_a & = & \beta A_u \;-\;\alpha e^{i\phi}A_v \label{baseqcl1}\\
A_b & = & \alpha A_u \;+\;\beta e^{i\phi}A_v \label{baseqcl2}
\eeqr

with $\alpha$, $\beta$ positive real numbers such that
$\alpha^2+\beta^2=1$. The correlation term reads

\beqr |\langle\daa\dab\rangle|^2\;=\;\alpha^2\beta^2 [\;
\langle\dau^2\rangle^2+\langle\dav^2\rangle^2-
2\langle\dau^2\rangle\langle\dav^2\rangle\cos2\phi\;]\label{correl-ab}\eeqr

and is maximal for $\phi=\pi/2\;[\pi]$ and
$\alpha=\beta=1/\sqrt{2}$. The maximally entangled modes are then
the circularly polarized modes with respect to modes $u,v$:

\beqr A_{a^*} & = & \frac{1}{\sqrt{2}} \;(\;A_u \;-\;iA_v\;) \label{basecorrele1}\\
 A_{b^*} & = & \frac{1}{\sqrt{2}}\;(\; A_u \;+\;iA_v\;) \label{basecorrele2} \eeqr

which satisfy

\beqr \nonumber|\langle\delta A_{a^*}\delta
A_{b^*}\rangle|=\max_{a,b}|\langle\daa\dab\rangle|=\frac{1}{2}
[\;\langle\dau^2\rangle+\langle\dav^2\rangle\;]\eeqr

Plugging this result in (\ref{valeurduan}) and using
(\ref{basecorrele1}-\ref{basecorrele2}), we obtain the maximal
entanglement value as the sum of the minimal noise of the
"uncorrelated" modes

\beqr
\mathcal{I}_{a^*,b^*}\;\equiv\;\min_{a,b}\iab\;=\;\langle\delta
X_u^2\rangle_{min}\;+\;\langle\delta
X_v^2\rangle_{min}\label{bestcorrel}\eeqr

This is the main result of this Section: \textit{the best
entanglement in the system is found between the circularly
polarized modes in the $(u,v)$ basis, and it is equal to the sum
of the $u,v$ mode minimal noises}. This result stresses the link
between entanglement and squeezing; associating to each
polarization basis $(a,b)$ the sum of the minimal noises

\beqr \Sigma_{a,b}&\equiv &\langle\delta
X_a^2\rangle_{min}+\langle\delta X_b^2\rangle_{min}\\
\nonumber&=&\langle\daa^{\dagg}\daa+\daa\daa^{\dagg}+\dab^{\dagg}\dab+\dab\dab^{\dagg}\rangle-
2[\;|\langle\daa^2\rangle|+|\langle\dab^2\rangle|\;]
\label{sommeminimal} \eeqr

Eq. (\ref{bestcorrel}) clearly shows that looking for maximal
entanglement is equivalent to looking for maximal squeezing
produced by the system

\beqr \mathcal{I}_{a^*,b^*}=\min_{a,b}\left[\langle\delta
X_a^2\rangle_{min}+\langle\delta X_b^2\rangle_{min}\right]\equiv
\Sigma_{min}\eeqr

\subsection{Correlations in the Poincar\'{e}
sphere}\label{poincaresphere}

\begin{figure}[h]
  \centering
  \includegraphics[width=15cm]{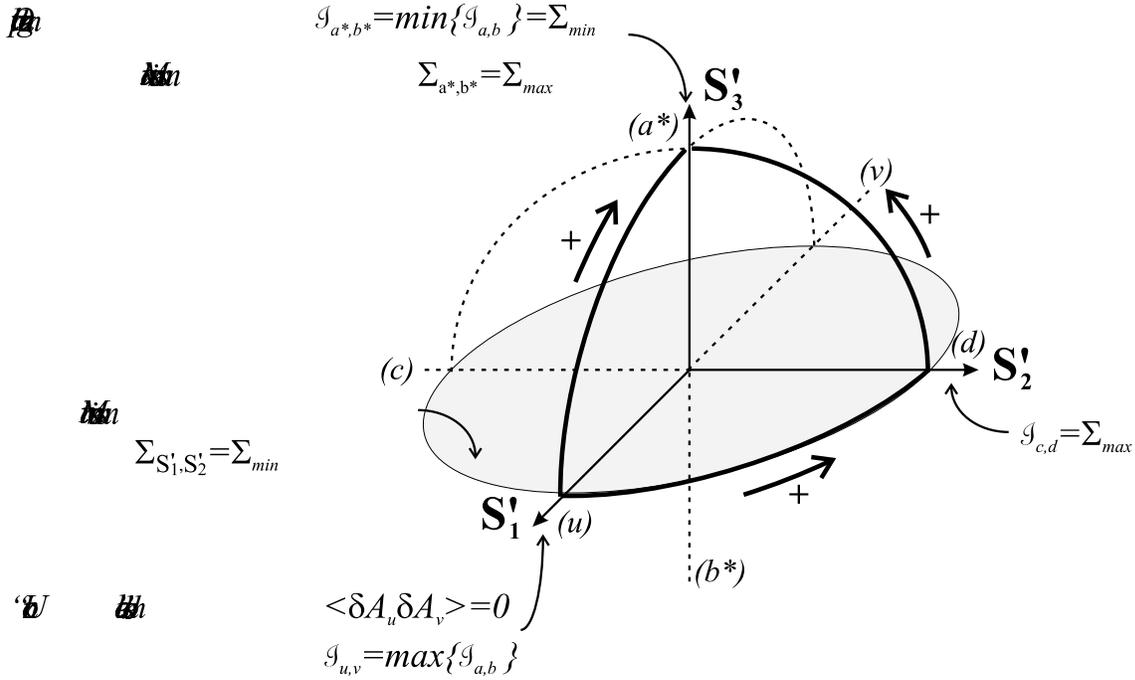}
  \caption{Quantum properties of the beam in the Poincar\'{e} sphere.
  The arrows with sign "+" correspond to increasing correlations.}\label{spherecorrel}
\end{figure}

A standard representation for the polarization state of light is
provided by the Poincar\'{e} sphere \cite{korolkova}, which relies
on the Stokes parameters \ct{huard}. Given the special role played
by the "uncorrelated" basis, we define the Stokes parameters from
the $u,v$ modes

\beqr S'_{0}&=&A^*_u A_u+ A^*_v A_v\hspace{1cm}
S'_{1}=A^*_u A_u- A^*_v A_v\nonumber\\
S'_{2}&=&A^*_u A_v+ A^*_v A_u\hspace{1cm} S'_{3}=i(A^*_v A_u-A^*_u
A_v)\nonumber\eeqr

and we study the evolution of entanglement and squeezing when the
polarization basis is rotated, that is, when the polarization
state vector moves along the Poincar\'{e} sphere. In the general
case the correlation properties of the system can be summarized as
follows (see Appendix B for the demonstration of these results):

\begin{enumerate}

\item Along the "uncorrelated" modes axis $S'_1$, $\iab$ is maximal by construction ($\langle \dau\dav\rangle
=0$) and these modes are never entangled

\beqr \nonumber\mathcal{I}_{u,v}&=&\max_{a,b}\iab=\langle
\dau^{\dagg}\dau+\dau\dau^{\dagg}+\dav^{\dagg}\dav+\dav\dav^{\dagg}\rangle\\
\nonumber &=&\langle \delta X_u^2\rangle+\langle \delta
Y_u^2\rangle+\langle \delta X_v^2\rangle+\langle \delta
Y_v^2\rangle\geq 2\eeqr

These modes are characterized by the fact that
$\mathcal{I}_{u,v}(\theta)$ is independent of $\theta$. The least
noisy quadratures are the same and the noise reduction is maximal

\beqr \nonumber\Sigma_{u,v}=\Sigma_{min}=\langle\delta
X_u^2\rangle_{min}+\langle\delta X_v^2\rangle_{min}\eeqr

\item In the equatorial plane $(S'_1,S'_2)$, corresponding to the
linearly polarized modes with respect to $u,v$, the noise
reduction is also maximal and equal to $\Sigma_{min}$. However,
the entanglement is not constant: the best entanglement is
obtained along the $S'_2$ axis (modes at $45^{\circ}$ to the $u,v$
modes) and its value is equal to the weakest noise reduction
$\Sigma_{max}$

\beqr
\nonumber\mathcal{I}_{c,d}=\Sigma_{max}=\min\left[\langle\delta
X_u^2\rangle_{min}+\langle\delta
X_v^2\rangle_{max}\;,\;\langle\delta
X_u^2\rangle_{max}+\langle\delta X_v^2\rangle_{min}\right]\eeqr

\item Along the $S'_3$ axis are the maximally entangled modes
$a^*,b^*$, for which the entanglement is maximal (and equal to the
best noise reduction value)

\beqr\nonumber
\mathcal{I}_{a^*,b^*}=\min_{a,b}\iab=\Sigma_{min}=\langle\delta
X_u^2\rangle_{min}+\langle\delta X_v^2\rangle_{min}\eeqr

and for which the excess noise is the largest

\beqr \nonumber\Sigma_{a^*,b^*}=\langle\delta
X_{a^*}^2\rangle_{min}+\langle\delta
X_{b^*}^2\rangle_{min}=\Sigma_{max}\eeqr

\end{enumerate}

A graphical representation of these results is given in Fig.
\ref{spherecorrel}: at the poles there is maximal entanglement
($\mathcal{I}_{a^*,b^*}=\Sigma_{min}$) and the worst noise
reduction ($\Sigma_{a^*,b^*}=\Sigma_{max}$). In the equatorial
plane the noise reduction is optimal
($\Sigma_{S'_1,S'_2}=\Sigma_{min}$). Note that the $c,d$ modes at
"$\pm 45^{\circ}$" may be entangled while having optimal noise
reduction.

\subsection{Interpretation}

\begin{figure}[h]
  \centering
  \includegraphics[width=12cm]{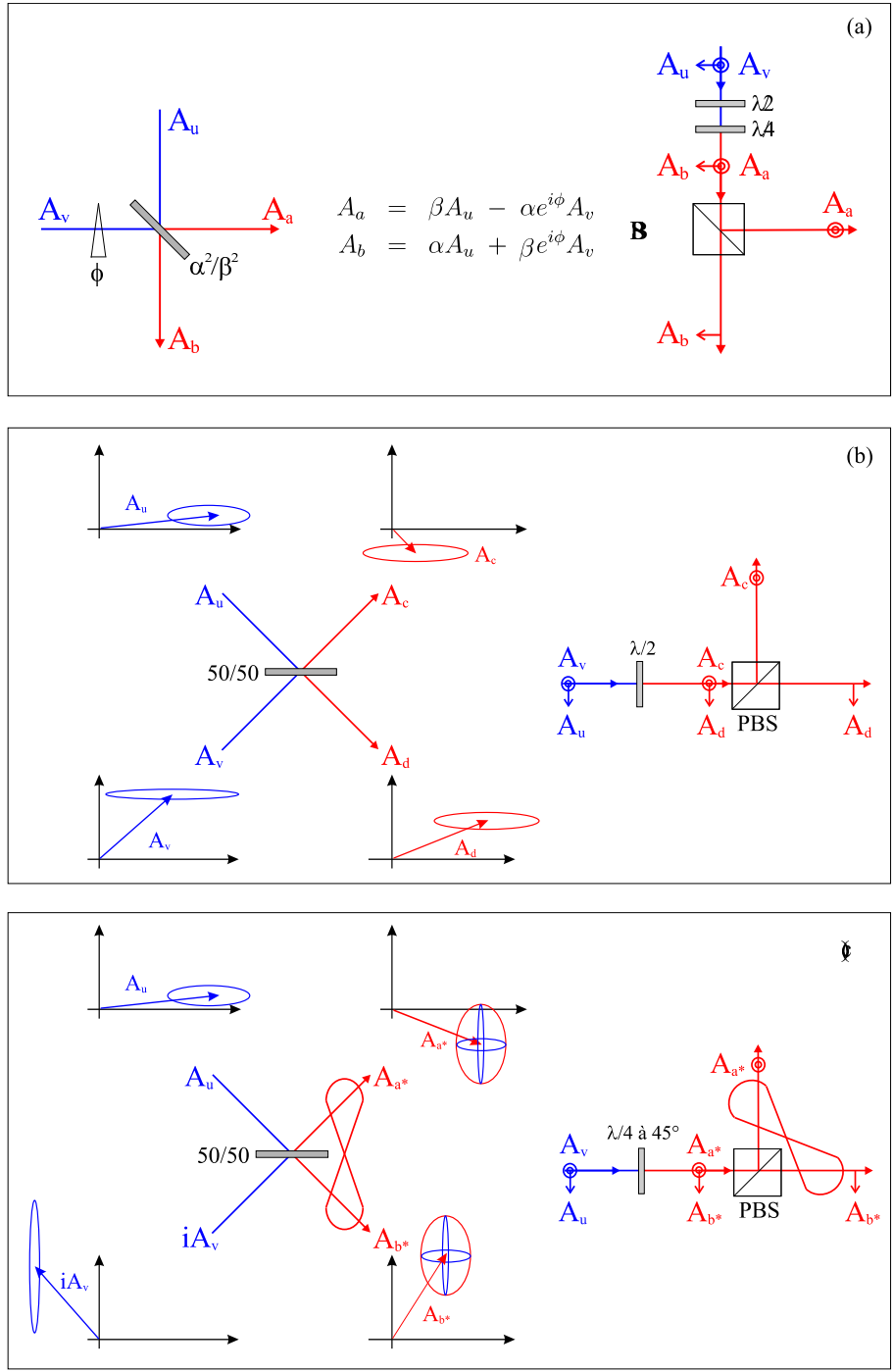}
  \caption{Interpretation of the correlations. (a) Equivalence between
  the transformations corresponding to the transmission
  by a beamsplitter (left) and a polarization basis rotation (right).
  (b) Squeezing conservation for the linearly polarized modes when there
  is no dephasing between the squeezed quadratures. (c) Generation of
  entangled modes via the interference of two modes squeezed for
  orthogonal quadratures.}\label{interpretcorrel}
\end{figure}

An interpretation of the previous results can be given in
connection with typical experiments in which entanglement is
obtained by having independent beams interfere with each other
\cite{bowen02prl,glockl}. Indeed, going from the ($u,v$) to the
($a,b$) basis with parameters $\alpha$, $\beta$ and $\phi$ [Eqs.
(\ref{baseqcl1},\ref{baseqcl2})] is equivalent to combining the
$u,v$ modes on a beamsplitter with transmission $T=\beta^2$, the
$v$ mode being dephased by $\phi$ [see Fig.
\ref{interpretcorrel}(a)]. Although, again, the $u,v$ modes are
not completely uncorrelated, we can interpret the general results
enunciated above as the result of an interference between two
independent beams. This configuration is the one typically used to
generate entangled beams: two squeezed beams on the same
quadrature are produced separately - for instance with two OPAs
\ct{bowen02prl,bowen03prl}, or by using the Kerr-type
non-linearity of optical fibers \ct{glockl} -
and then combined on a beamsplitter.\\
The $u,v$ modes are squeezed for the same quadrature $Y$, so that,
in case of a zero-dephasing before the beamsplitter, the squeezed
and noisy quadratures do not mix and the noise reduction sum is
conserved [Fig.\ref{interpretcorrel}(b)]. The outgoing beams are
linearly polarized with respect to the incoming ones ($\phi=0$)
and we retrieve the property (ii)

\beqr \nonumber\Sigma_{S'_1,S'_2}=\langle\delta
X_u^2\rangle_{min}+\langle\delta
X_v^2\rangle_{min}=\Sigma_{min}\eeqr

If the dephasing before mixing is now equal to $\pi/2$, the
outgoing beams have excess noise on both quadratures [see Fig.
\ref{interpretcorrel}(c)]. However, these noises are correlated
and, obviously, all the more so for a 50/50 beamsplitter, since it
maximally mixes the $u,v$ modes. This transformation is equivalent
to going from the linearly polarized basis ($u,v$) to the
circularly polarized basis ($a^*,b^*$), which is naturally the
correlated basis

\beqr \nonumber &&\frac{1}{2}\langle
\delta(X_{a^*}+X_{b^*})^2(\theta)\rangle=\langle \delta
X_u^2(\theta)\rangle<1 \hspace{0.3cm}for\hspace{0.2cm}\theta=\frac{\pi}{2}\\
\nonumber
&&\frac{1}{2}\langle\delta(Y_{a^*}-Y_{b^*})^2(\theta)\rangle=\langle
\delta X_v^2(\theta)\rangle<1
\hspace{0.3cm}for\hspace{0.2cm}\theta=\frac{\pi}{2}\\
\nonumber Thus\hspace{0.5cm}&&\mathcal{I}_{a^*,b^*}= \langle
\delta X_u^2\rangle_{min}+\langle \delta X_v^2\rangle_{min}<2\eeqr

When the $u,v$ modes are symmetrical (same noise properties), it
can be shown that they are completely independent (see Sec.
\ref{quadentang} for an example). The $a^*,b^*$ modes fluctuations
are then the same for all quadratures.\\

Before applying these results to our experiment, we would like to
emphasize that this analogy between the quantum properties of any
system and those produced via the mixing of two independent beams
provides us with a simple interpretation of the main results. This
analysis is of particular interest to the study of systems for
which the correlations are simultaneously produced inside a "black
box". One has to find the "uncorrelated" modes $u,v$, an
experimental signature being that the entanglement value
$\mathcal{I}_{u,v}(\theta)$ does not depend on $\theta$. Once this
basis is obtained, one can apply the previous formalism or,
equivalently, think in terms of independent beams combination.
This method allows, for instance, to determine the maximal
correlations produced by an OPO inside which are inserted
birefringent elements \ct{fabre}.

\section{Quadrature entanglement}\label{quadentang}

In our experiment \ct{josse2}, the black box consists in an
optical cavity containing a cloud of cold cesium atoms into which
is sent an off-resonant light beam. In this Section we first
present the principle of a detection scheme allowing for a direct
measurement of the quadrature entanglement given by
(\ref{critere}). We then study the entanglement generated in the
case of a linear incident polarization, which is qualitatively
different from the circular polarization case.

\subsection{Entanglement measurement principle}

\begin{figure}[h]
  \centering
  \includegraphics[width=11cm]{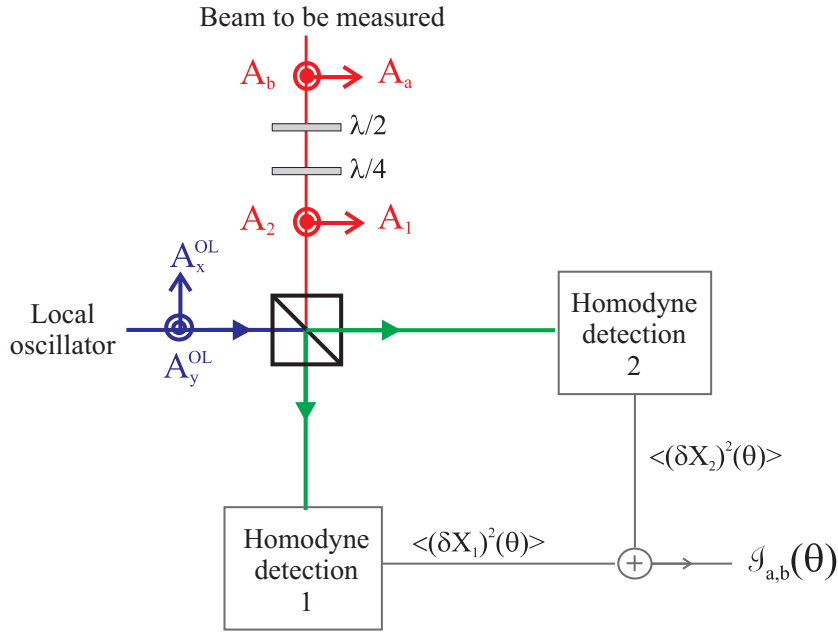}
  \caption{Schematic of the $a,b$ modes entanglement measurement.}\label{principeDuan2}
\end{figure}

In order to measure the entanglement given by (\ref{critere})
between two orthogonally polarized modes $a$ and $b$, we can
reexpress $\iabt$ as the sum of the noises of circularly polarized
modes $1,2$ with respect to $a,b$

\beqr \nonumber\iabt &= &\langle\delta X_1^2
(\theta)\rangle+\langle\delta X_2^2 (\theta)\rangle\\
with\hspace{0.3cm} A_1 & =&  \frac{1}{\sqrt{2}}(A_a\;+\;A_b)\nonumber\\
\hspace{1.2cm}A_2  &= &
\frac{i}{\sqrt{2}}(A_a\;-\;A_b)\nonumber\eeqr

As represented in Fig. \ref{principeDuan2}, the modes $1,2$ are
straightforwardly obtained from the given $a,b$ modes with a
half-wave and a quarter-wave plate. They are then mixed with a
strong coherent local oscillator (LO) on a polarizing beamsplitter
and sent to two balanced homodyne detections. We thus
simultaneously measure the spectral noise densities $\langle\delta
X_{1}^2(\theta)\rangle$ and $\langle\delta X_{2}^2(\theta)\rangle$
at a given analysis frequency, the sum of which directly gives
$\iabt$. This value oscillates when the LO phase is varied in
time. Note that, unlike usual detection schemes \ct{bowen02prl}
involving two successive measurements, this method is based on one
simultaneous measurement.

\subsection{Case of a linear incident polarization}

\begin{figure}[h]
  \centering
  \includegraphics[width=15cm]{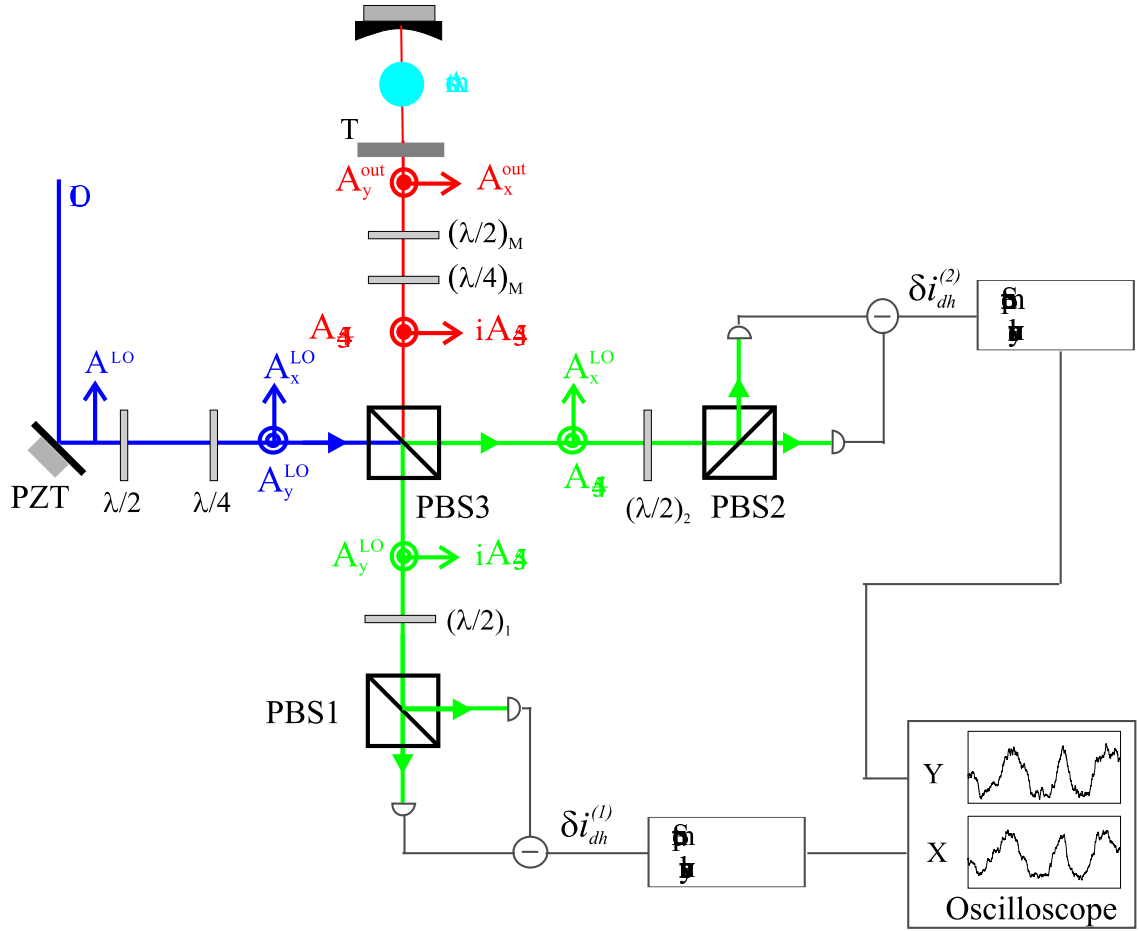}
  \caption{Experimental set-up.}\label{principeDuan}
\end{figure}

\begin{figure}[h]
  \centering
  \includegraphics[width=5cm]{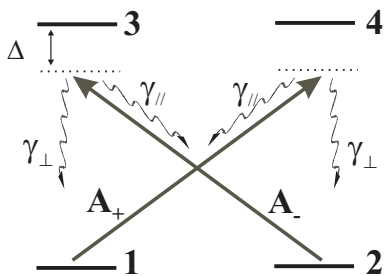}
  \caption{Atomic level structure considered: X-like configuration.}\label{X}
\end{figure}

In the system considered in \ct{josse3}, an $x$-polarized beam
interacts with a cloud of cold cesium atoms in an optical cavity.
The experimental set-up is shown in Fig. \ref{principeDuan}. The
cavity coupling mirror has a transmission coefficient of 10\%, the
rear mirror is highly reflecting. We probe the atoms with a
linearly polarized laser beam detuned by about 50 MHz in the red
of the 6S$_{1/2}$, F=4 to 6P$_{3/2}$, F=5 transition. The optical
power of the probe beam ranges from 5 to 15 $\mu$W. After exiting
the cavity, both the mean field mode $A_x$ and the orthogonally
polarized vacuum mode $A_y$ are squeezed for frequencies ranging
between 3 and 12 MHz. An interpretation of these results
\ct{josse1} can be provided by modelling the complicated
6S$_{1/2}$, F=4 to 6P$_{3/2}$, F=5 transition by an X-like
four-level atomic structure (Fig. \ref{X}).

When the two transitions are symmetrically saturated, the atoms
behave as a Kerr-like medium for the circular components

\beqr A_{\pm}=\frac{1}{\sqrt{2}}(A_x\;\mp \;iA_y)\label{Apm}\eeqr

both of which are squeezed \ct{josse1}. Because of the symmetry of
the system, they are obviously squeezed for the same quadrature.
From the viewpoint of linear polarizations the $x,y$ modes are
also squeezed due to cross-Kerr effect,
but for \textit{orthogonal} quadratures \ct{josse1,josse2}.\\
Indeed, from (\ref{Apm}), one derives the following expressions

\beqr \langle\dax\dai\rangle
&=&\frac{i}{\sqrt{2}}(\langle\dacp^2\rangle -
\langle\dacm^2\rangle + \langle\dacp\dacm^{\dagg}\rangle
-\langle\dacp^{\dagg}\dacm\rangle)=0\label{correl-xy1}\\
 \langle\dax\dai^{\dagg}\rangle &=&\frac{-i}{\sqrt{2}}(\langle\dacp\dacp^{\dagg}\rangle -
\langle\dacm\dacm^{\dagg}\rangle +
\langle\dacp\dacm^{\dagg}\rangle
-\langle\dacp^{\dagg}\dacm\rangle)=0
\label{correl-xy2}\\
\langle\dax^2\rangle &=&\frac{1}{\sqrt{2}}(\langle\dacp^2\rangle +
\langle\dacm^2\rangle -
2\langle\dacp\dacm\rangle )\label{quadx}\\
 \langle\dai^2\rangle &=&-\frac{1}{\sqrt{2}}(\langle\dacp^2\rangle + \langle\dacm^2\rangle +
2\langle\dacp\dacm\rangle )\label{quady}\eeqr

Eqs. (\ref{correl-xy1}-\ref{correl-xy2}) show that the $x,y$ modes
are completely independent. We then measure
$\mathcal{I}_{x,y}(\theta)$ following the previous procedure:
$A_{+45}$ and $iA_{-45}$ are sent to the homodyne detections (Fig.
\ref{principeDuan}), yielding the quantity

\beqr \mathcal{I}_{x,y}(\theta) &= &\langle\delta X_{+45}^2
(\theta)\rangle\;+\;\langle\delta X_{i(-45)}^2(\theta)\rangle
\label{mesureduan-xy}\\
with\hspace{0.3cm} A_{+45} & =&  \frac{1}{\sqrt{2}}(A_x\;+\;A_y)\nonumber\\
\hspace{1.1cm} iA_{-45}  &=&
\frac{i}{\sqrt{2}}(A_x\;-\;A_y)\nonumber \eeqr

\begin{figure}[h]
  \centering
  \includegraphics[width=10cm]{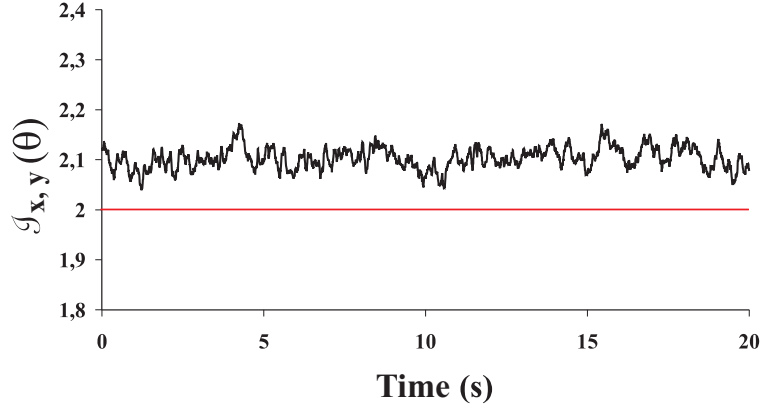}
  \caption{$\mathcal{I}_{x,y}(\theta)$ when $\theta$ is varied in time, at $5MHz$.}\label{independants-xy}
\end{figure}

We verify in Fig. \ref{independants-xy} that this quantity is
effectively independent of $\theta$, ensuring that the $x,y$ modes
are "uncorrelated" in the
sense of the inseparability criterion.\\
Bearing in mind that the $u,v$ mode noises must be minimal for the
same quadrature, we look at (\ref{quadx}-\ref{quady}) to find the
relative orientation of the minimal quadratures of the $x,y$
modes. These minimal quadratures are \textit{a priori} different
and depend on the correlation term $\langle \delta A_+\delta
A_-\rangle$ between the circularly polarized modes. Physically,
these modes are correlated by optical pumping processes between
Zeeman sublevels. However, we place ourselves in the "high
frequency" limit for which the analysis frequency (a few MHz) is
much higher than the optical pumping rate (a few hundreds of kHz).
The modes $A_{\pm}$ are then uncorrelated ($\langle \delta
A_+\delta A_-\rangle\simeq 0$), and squeezed for the same
quadratures. Using (\ref{quadx}-\ref{quady}) and the fact that

\begin{figure}[h]
  \centering
  \includegraphics[width=10cm]{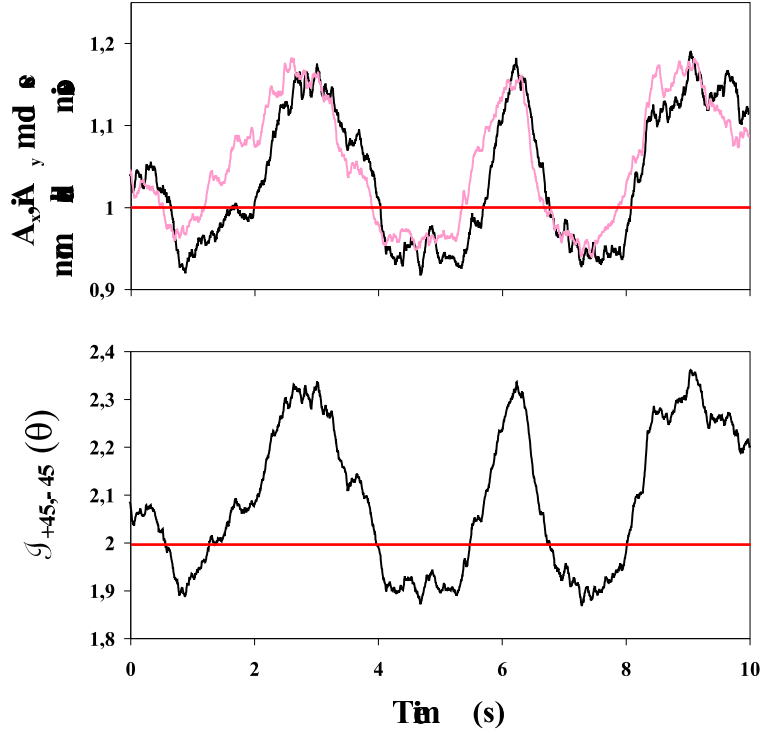}
  \caption{(a) Noise spectra of $A_x$ and $iA_y$ when the LO phase is varied in time. The analysis frequency is $5MHz$.
  (b) Corresponding value of $\mathcal{I}_{+45,-45}(\theta)$.}\label{bonduan}
\end{figure}

\beqr\nonumber \langle \delta X_{\alpha}^2\rangle=\langle \delta
A_{\alpha}\delta A_{\alpha}^{\dagger}+\delta
A_{\alpha}^{\dagger}\delta A_{\alpha}\rangle+2\langle\delta
A_{\alpha}^2\rangle \cos2\theta,\hspace{0.5cm}(\alpha=u,v)\eeqr

we deduce that the $x,y$ modes are squeezed for
\textit{orthogonal} quadratures. These properties were verified
both theoretically \ct{josse1} and experimentally \ct{josse2} [see
also Fig. \ref{bonduan}(a)]. In the high frequency limit, one has
thus to dephase one mode by $\pi/2$; we choose for "uncorrelated"
basis $A_u=A_x$ and $A_v=iA_y$.\\
The maximally entangled modes are then the modes at $45^{\circ}$
to the $x,y$ basis

\beqr A_{a^*}&=&\frac{1}{\sqrt{2}}(A_u-iA_v)=\frac{1}{\sqrt{2}}(A_x+A_y)\equiv A_{+45}\\
A_{b^*}&=&\frac{1}{\sqrt{2}}(A_u+iA_v)=\frac{1}{\sqrt{2}}(A_x-A_y)\equiv
A_{-45}\eeqr

and their entanglement is measured by summing the minimal noises
of $A_x$ and $iA_y$

\beqr \mathcal{I}_{+45,-45} &= &\langle\delta X_{x}^2
\rangle_{min}\;+\;\langle\delta X_{iy}^2\rangle_{min}\simeq 1.9
\label{mesureduan-45} \eeqr

The results are reproduced in Fig. \ref{bonduan} for an analysis
frequency of $5MHz$: the two modes have indeed the same spectrum,
and, when summed, the minimal value of
$\mathcal{I}_{+45,-45}(\theta)$ is below 2, demonstrating
entanglement. Equivalently, one could have set $A_u=A_+$ and
$A_v=A_-$, since these modes are uncorrelated and symmetrical. To
obtain the entangled modes, one has to dephase them by $\pi/2$ and
combine them on a beamsplitter, yielding again the $A_{\pm 45}$
modes

\beqr \nonumber A_{+45} &\equiv &\frac{1}{\sqrt{2}}(A_x+A_y)=-\frac{e^{i\frac{\pi}{4}}}{\sqrt{2}}(A_+ + iA_-)\\
\nonumber A_{-45} &\equiv & \frac{1}{\sqrt{2}}(A_x-A_y)=
-\frac{e^{-i\frac{\pi}{4}}}{\sqrt{2}}(A_+ - iA_-)\eeqr

\begin{figure}[h]
  \centering
  \includegraphics[width=8cm]{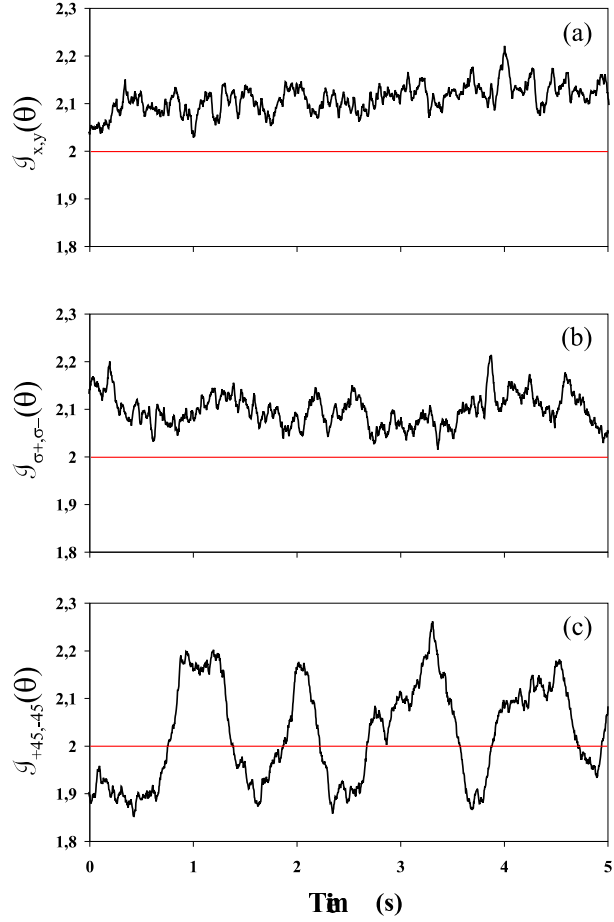}
  \caption{Entanglement for the canonical polarization basis:
  $(x,y)$ (a), $(\sigma_+,\sigma_-)$ (b), $(+45,-45)$ (c),
  for an analysis frequency of $5MHz$.}\label{intrication-modes}
\end{figure}

Let us now represent these results in the Poincar\'{e} sphere. In
contrast with the previous Section, we define the Stokes
parameters in a more usual fashion from the linearly polarized
modes ($x,y$) basis

\beqr S_{0}&=&A^*_x A_x+ A^*_y A_y\hspace{0.5cm}
S_{1}=A^*_x A_x- A^*_y A_y\nonumber\\
S_{2}&=&A^*_x A_y+ A^*_y A_x\hspace{0.5cm} S_{3}=i(A^*_y A_x-A^*_x
A_y)\nonumber\eeqr

With the relations $A_u=A_x,\hspace{0.2cm}A_v=iA_y$, it is easy to
see that the new Poincar\'{e} sphere is obtained from the previous
one via a rotation by $\pi/2$ around the $S_1$ axis. If we measure
the entanglement in the ($\sigma+,\sigma-$) basis

\beqr \nonumber\mathcal{I}_{\sigma+,\sigma-}(\theta) &=
&\langle\delta X_{x}^2 (\theta)\rangle\; +\;\langle\delta
X_{y}^2(\theta)\rangle\eeqr

and represent the results in Fig. \ref{intrication-modes}, one
sees that the basis in the ($S_1,S_3$) plane are uncorrelated, as
expected, whereas maximal
entanglement is found for the $\pm 45^{\circ}$ modes.\\
We also measured the noise spectra of each modes in the three
basis and checked that the squeezing is maximal and identical for
all modes in the ($S_1,S_3$) plane

\begin{figure}[h]
  \centering
  \includegraphics[width=15cm]{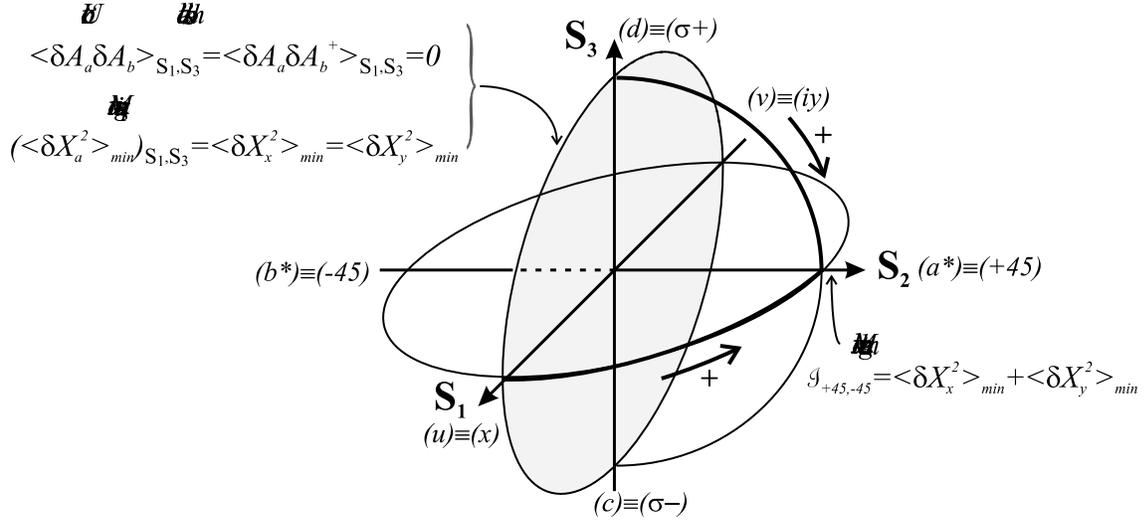}
  \caption{Quantum properties of the beam in the Poincar\'{e} sphere
at high frequency, in the linear polarization
case.}\label{spherecorrel2}
\end{figure}

\beqr\nonumber\langle\delta X_x^2\rangle_{min}=\langle\delta
X_y^2\rangle_{min}=\langle\delta
X_{\sigma+}^2\rangle_{min}=\langle\delta
X_{\sigma-}^2\rangle_{min}\simeq 0.95\eeqr

and that the entangled modes $A_{\pm 45}$ have almost isotropic
fluctuations in the Fresnel diagram, as well as identical
spectra because of the independence of the $x,y$ modes.\\
One can summarize these results in the Poincar\'{e} sphere
represented in Fig. \ref{spherecorrel2}.

\subsubsection{Frequency dependence}

\begin{figure}[h]
\centering
\includegraphics[width=15cm]{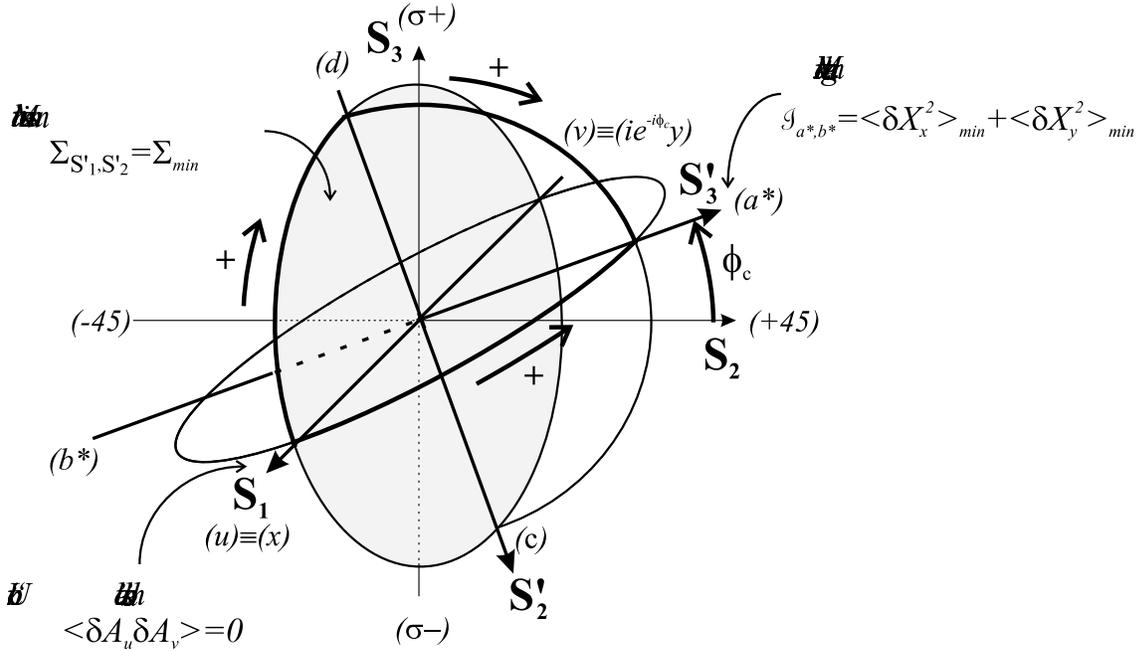}
\caption{Quantum properties of the beam in the Poincar\'{e} sphere
\textit{at lower frequency}, in the linear polarization case.}
 \label{spherecorrel3}
\end{figure}

The situation is a little bit more complicated at lower
frequencies. The correlations between the circularly polarized
components must be taken into account for frequencies lower than
the optical pumping rate. The $\sigma\pm$ modes are still
symmetrical and the $x,y$ modes still uncorrelated. However, the
minimal noise quadratures are rotated [Eqs.
(\ref{quadx}-\ref{quady})] and the $x,y$ modes are no longer
squeezed for orthogonal quadratures. To retrieve the
"uncorrelated" basis, one mode must be dephased, say $y$:

\beqr \nonumber A_u=A_x\;\;\;\;and\;\;\;\;A_v=ie^{-i\phi_C}A_y
\eeqr \beqr\nonumber where \;\;\;\;\;\tan
\phi_C=\frac{2|\langle\dacp\dacm\rangle|\sin (\phi_{2}-\phi_{1
})}{\langle\dacp^2\rangle + \langle\dacm^2\rangle +
2|\langle\dacp\dacm\rangle|\cos (\phi_{1}-\phi_{2}) } \eeqr

In the high frequency limit,
$|\langle\dacp\dacm\rangle|\rightarrow 0$ and, therefore
$\phi_C\rightarrow 0$: we retrieve the previous ($u,v$) basis.
This dephasing for the $y$ mode is equivalent to a rotation of the
Poincar\'{e} sphere by an angle $\phi_C$ around the $S_1$ axis
[Fig. \ref{spherecorrel3}]. The maximally entangled modes are then

\beqr
A_{a^*}&=&\frac{1}{\sqrt{2}}(A_u-iA_v)=\frac{1}{\sqrt{2}}(A_x+e^{-i\phi_{C}}A_y)\label{Arot1}\\
A_{b^*}&=&\frac{1}{\sqrt{2}}(A_u+iA_v)=\frac{1}{\sqrt{2}}(A_x-e^{-i\phi_{C}}A_y)\label{Arot2}\eeqr

and their entanglement is still given by (\ref{mesureduan-45})

\beqr \nonumber\mathcal{I}_{a^*,b^*}=\langle\delta
X_x^2\rangle_{min}+\langle\delta X_y^2\rangle_{min} \eeqr

\begin{figure}[h]
\centering
\includegraphics[width=12cm]{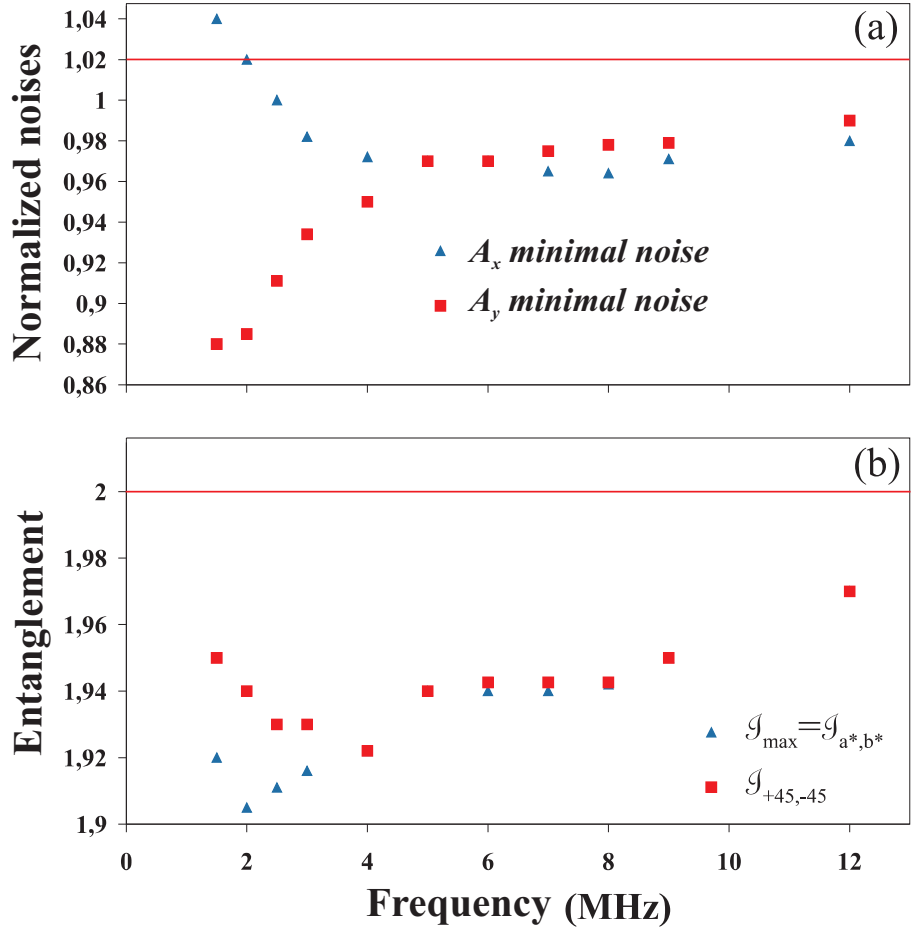}
\caption{(a) $x,y$ modes minimum noises versus frequency. (b)
Entanglement for the $\pm 45^{\circ}$-polarized modes
($\mathcal{I}_{+45,-45}$) compared to the maximal entanglement
($\mathcal{I}_{a^*,b^*}$).} \label{duan-vs-frequence}
\end{figure}

In Fig. \ref{duan-vs-frequence} are plotted the $x,y$ modes
squeezing versus frequency, as well as $\mathcal{I}_{+45,-45}$ and
the optimal entanglement $\mathcal{I}_{a^*,b^*}$. At low frequency
the squeezing improves for the vacuum mode $A_y$, but degrades for
the mean field mode $x$, so that the entanglement actually
decreases at low frequency. In Fig. \ref{duan-vs-frequence} we
also report the value of $\mathcal{I}_{+45,-45}$. One sees that it
equals the optimal entanglement in the high frequency limit, but,
for lower frequencies, the two values differ, confirming that the
maximally entangled modes are no longer $A_{\pm 45}$, but given by
(\ref{Arot1}-\ref{Arot2}).

\subsection{Case of a circular incident polarization}

In this Section, we illustrate the differences between a
\textit{two squeezed mode} system and a \textit{single squeezed
mode} system. We show that the entanglement produced in the latter
is qualitatively different from the former, even if the amount of
correlations is the same. Whereas the two squeezed modes situation
corresponds to the case of a linear polarization, the single
squeezed mode situation appears when the polarization is circular.
Indeed, our system may exhibit \textit{polarization switching}:
the intracavity polarization may become circular under some
conditions \ct{josse1}. In this case, the atoms only interact with
one mode - say $\sigma_+$, which may also be squeezed because of
Kerr effect. Yet, the situation is very different from the
previous one. One can set $A_u=A_+$ and $A_v=A_-$, the former
being squeezed, the latter being a coherent vacuum. The vacuum
fluctuations are isotropic and, therefore, the properties of the
($u,v$) basis must remain unchanged when the $v$ mode is dephased.
In other words, the Poincar\'{e} sphere must be invariant under
rotations around $S_1$. Moreover, since $\langle \dav^2\rangle=0$,
the minimal noise sum is uniform on the whole sphere

\beqr\nonumber \Sigma_{min}=\Sigma_{max}= \langle\delta
X_u^2\rangle_{min}+1\eeqr

The entanglement is maximal and constant in the ($S_2,S_3$) plane

\beqr\nonumber \mathcal{I}_{S_2,S_3}= \min_{a,b} \iab=
\langle\delta X_u^2\rangle_{min}+1\eeqr

In this plane all modes have identical noise

\beqr \langle \delta X_a^2(\theta)\rangle=\frac{1}{2}(1+\langle
\delta X_u^2(\theta)\rangle)\label{moitie}\eeqr

This result is easily understandable if compared to the
transmission of a squeezed beam by a beamsplitter. It is well
known that part of the incident beam squeezing is lost to the
reflected beam and leaks into the environment.\\
We measured the noise spectra for all basis. We observed squeezing
in the $\sigma_+$ component, the $\sigma_-$ component being at the
shot noise level. We checked that the spectra of the $x$, $iy$,
$+45$, $i(-45)$ modes were all identical and squeezed by half the
amount of the $\sigma_+$ component
squeezing, consistently with (\ref{moitie}).\\

\begin{figure}[h]
\centering\includegraphics[width=8cm]{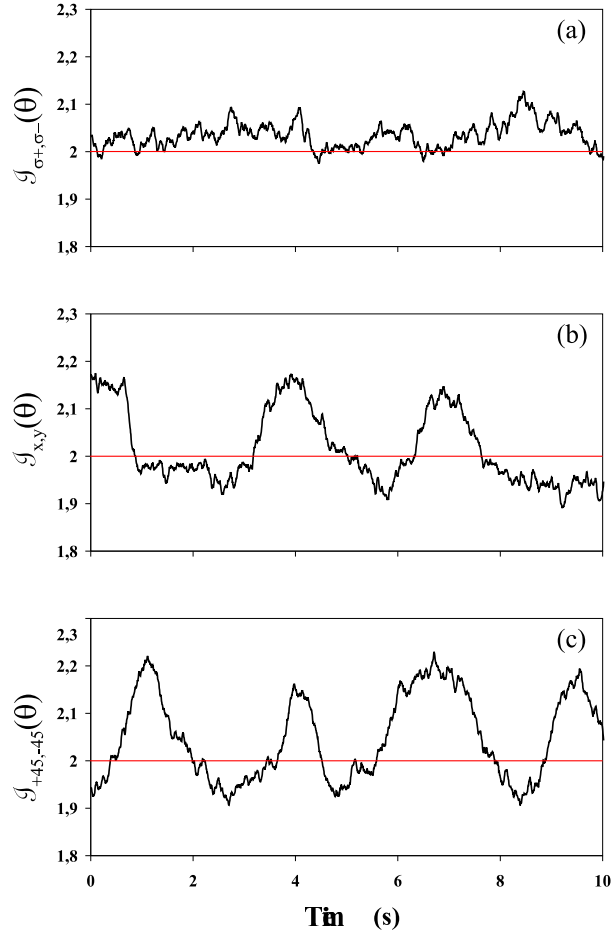} \caption{Entanglement
in different polarization basis, in the circular polarization
case: (a) $\mathcal{I}_{x,y}(\theta)$, (b)
$\mathcal{I}_{\sigma+,\sigma-}(\theta)$, (c)
$\mathcal{I}_{+45,-45}(\theta)$.}
\label{intrication-modes-circulaire}
\end{figure}

Although the $x,y$ modes are both squeezed for orthogonal
quadratures, the difference with the previous case is that they
are now correlated. We verified this by measuring the entanglement
in each basis. The results, displayed in Fig.
\ref{intrication-modes-circulaire}, show that the circular
components are indeed uncorrelated, while the $x,y$ modes are
entangled, as well as the $\pm 45$ modes.

\section{Polarization entanglement}\label{polaentang}

Up to now, we have determined the correlation properties of the
beam and shown that non-separable states, namely the $\pm 45$
modes in the high frequency limit, were produced in our system. We
show in this Section that the quadrature entanglement demonstrated
previously can be mapped into a polarization basis, thus achieving
\textit{polarization entanglement}.

\subsection{Definition and scheme}

\begin{figure}[h]
\centering\includegraphics[width=10cm]{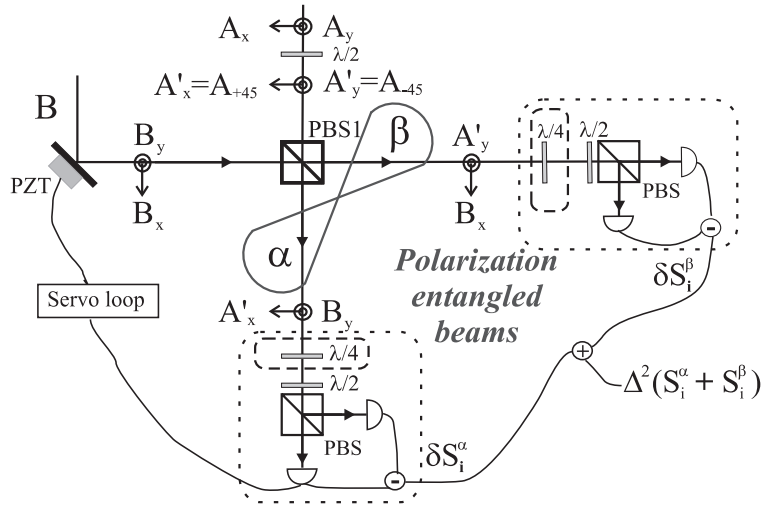}
\caption{Polarization-entangled beams generation set-up. Using the
quarter-wave plates or not allows for measuring either $S_{S_2}$
or $S_{S_3}$.} \label{intr-pola-schema}
\end{figure}

Let us consider two spatially separated polarization modes
$\alpha$ and $\beta$. To each one we associate a set of Stokes
parameters. One can extend the inseparability criterion for two
modes to a pair of these operators
\ct{josse3,korolkova,bowen02prl}. \textit{Polarization
entanglement} is achieved when

\beqr  \mathcal{I}_{\alpha,\beta}^{S} =\frac{1}{2} [\langle
\delta(S_2^{\alpha}+S_2^{\beta})^2\rangle+\langle\delta(S_3^{\alpha}+S_3^{\beta})^2\rangle]
 < |\langle[S_2^{\alpha}, S_3^{\alpha}]\rangle| +|\langle
[S_2^{\beta},S_3^{\beta}]\rangle|\label{Ialphabeta}\eeqr

Because of the cyclical commutation relations between the Stokes
operators, the criterion now depends on the polarization state of
the beams (here, $\langle S_1^{\alpha}\rangle$ and $\langle
S_1^{\beta}\rangle$). In our case, we use two quadrature entangled
modes $a$ and $b$, mix them on a polarizing beamsplitter with an
intense coherent beam $B$, polarized at $45^{\circ}$ [Fig.
\ref{intr-pola-schema}]. The resulting beams, $\alpha$ and
$\beta$, are composed with modes $A_a$ and $B_y$, and $A_b$ and
$B_x$ respectively. The Stokes parameters are

\beqr S_{1}^\alpha &=&A_a^{\dagger} A_a- B_y^{\dagger}
B_y\hspace{0.5cm}\;\;\;\;
S_{1}^\beta=B_x^{\dagger} B_x- A_b^{\dagger} A_b\nonumber\\
 S_{2}^\alpha &=&A_a B_y^{\dagger} + A_a^{\dagger} B_y\hspace{0.5cm}\;\;\;\;
S_{2}^\beta=A_b^{\dagger}B_x + A_bB_x^{\dagger} \nonumber\\
 S_{3}^\alpha &=&i(A_a B_y^{\dagger} - A_a^{\dagger} B_y)\hspace{0.5cm} S_{3}^\beta=i(A_b^{\dagger}B_x -
 A_bB_x^{\dagger})\nonumber\eeqr

Denoting by $\alpha_B$, $\alpha_a$, $\alpha_b$ and $\theta_B$ the
field amplitudes and the B-field phase, and assuming that the
B-field is much more intense than the A-field
($\alpha_B\gg\alpha_a,\alpha_b$), the two beams are orthogonally
polarized: $ \langle S_1^{\alpha}\rangle=-\langle
S_1^{\beta}\rangle=-\alpha_B^2$. The polarization entanglement
condition (\ref{Ialphabeta}) then reads

\beqr \mathcal{I}^S_{\alpha,\beta}<2\alpha_B^2\eeqr

On the other hand, the Stokes parameters fluctuations are
proportional to the $a,b$ modes quadratures

\beqr \nonumber\delta S_2^{\alpha} & = & \alpha_{B} \delta X_{a}
(\theta_B) ,\hspace{1cm}\;\;\delta S_2^{\beta}\; = \;\alpha_{B}
\delta
X_{b} (\theta_B)\\
\nonumber \delta S_3^{\alpha}& = &- \alpha_{B} \delta
Y_{a}(\theta_B),\hspace{0.9cm}\delta S_3^{\beta}\; =\; \alpha_{B}
\delta Y_{b}(\theta_B)\eeqr

The inseparability criterion is thus directly related to the
entanglement between the $a,b$ modes

\beq \iabs\;=\;\alpha_B^2 \;\iab (\theta_B) \eeq

If one locks the phase $\theta_B$ in order to obtain $\iab
(\theta_B)=\min_\theta \iabt\equiv\iab<2$, the beams are then
polarization entangled. The $S_2$ and $S_3$ Stokes parameters can
be measured using the right combination of plates and beamsplitter
\ct{korolkova}.

\subsection{Experimental results}

\begin{figure}[h]
\centering\includegraphics[width=10cm]{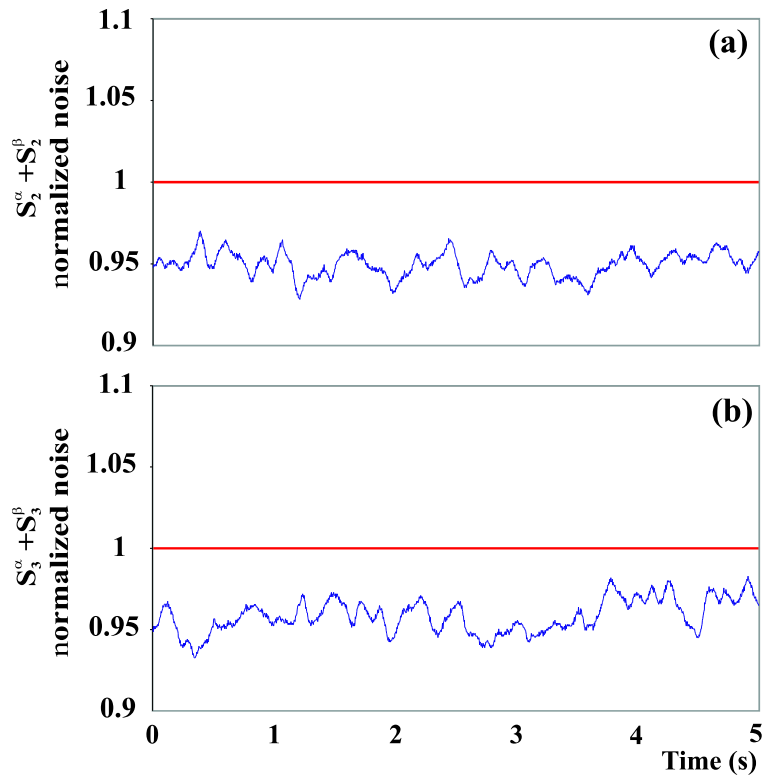} \caption{Normalized
noises of $S_2^{\alpha}+S_2^{\beta}$ (a) and
$S_3^{\alpha}+S_3^{\beta}$ (b), when the phase $\theta_B$ is
locked.} \label{intr-pola-mesure}
\end{figure}

We have shown that the maximally entangled modes were the $\pm 45$
modes at the output of the cavity. We thus insert after the cavity
a half-wave plate to set $A_a=A_{+45}$ and $A_b=A_{-45}$. We then
lock the phase $\theta_B$ and measure

\beqr &S_{S_2}&=\langle (\delta
X_{x}(\theta_B))^2\rangle\;\;\;,\;\;\; S_{S_3} = \langle(\delta
Y_{y}(\theta_B))^2\rangle\eeqr

The results are shown in Fig. \ref{intr-pola-mesure} and we
observe

\beqr\nonumber S_{S_2}\equiv \frac{1}{2\alpha_B^2}\langle
\delta(S_2^{\alpha}+S_2^{\beta})^2\rangle\simeq0.96<1\\
\nonumber S_{S_3}\equiv \frac{1}{2\alpha_B^2}\langle
\delta(S_3^{\alpha}+S_3^{\beta})^2\rangle\simeq0.96<1 \eeqr

so that

\beqr\nonumber \frac{\iabs}{\alpha_B^2}=S_{S_2}+S_{S_3}
=\mathcal{I}_{+45,i(-45)}\simeq1.92<2\eeqr

This value is consistent with the quadrature entanglement
measurement (\ref{mesureduan-45}).

\section{Conclusion}
Using the non-linearity of cold atoms, we have generated a
quadrature-entangled beam. The maximally entangled polarization
modes have been found following a general method to characterize
entanglement in a two mode system. We have stressed the
equivalence between our scheme and other continuous variable
entanglement experiments in which the correlations are created via
the mixing of independent beams. Moreover, a simple interpretation
of the quantum properties of such a system was given in the
Poincaré sphere. To demonstrate the entanglement we have performed
a direct measurement of the inseparability criterion \ct{duan}
using two homodyne detections. We have then achieved polarization
entanglement by mixing our quadrature entangled beam with an
intense coherent field. Experimental evidence of this entanglement
was given by the direct measurement of the Stokes operators noises
for each beam.

\appendix
\section{Uncorrelated basis existence}

Starting from a "correlated" polarization basis $(a,b)$
($\langle\daa\dab\rangle\neq 0$), we prove the existence of the
"uncorrelated" basis $(u,v)$, such that

\beqr \langle \dau\dav\rangle=0\label{uv}\eeqr

Using the following decomposition

\beqr A_u & = & \cos\Phi A_a \;-\;\sin\Phi e^{i\omega}A_b \nonumber\\
A_v & = & \sin\Phi A_a \;+\;\cos\Phi e^{i\omega}A_b \nonumber
\eeqr

with

\beqr e^{i\omega}=\frac{\mathcal{M}}{|\mathcal{M}|},\hspace{0.5cm}
\cos2\Phi=\frac{|\langle\daa^2\rangle|^2-|\langle\dab^2\rangle|^2}{\mathcal{N}},
\hspace{0.5cm}\sin2\Phi=\frac{-2|\mathcal{M}|}{\mathcal{N}}\nonumber\\
where\hspace{0.5cm}
\mathcal{M}=\langle\daa^2\rangle\langle\daa\dab\rangle^*+
\langle\dab^2\rangle^*\langle\daa\dab\rangle\nonumber\\
and\hspace{0.9cm}\mathcal{N}=\sqrt{4|\mathcal{M}|^2+[\;|\langle\daa^2\rangle|^2-|\langle\dab^2\rangle|^2\;]^2}\nonumber\eeqr

it is straightforward to see that the $u,v$ modes thus defined
satisfy (\ref{uv}).

\section{Correlations in the Poincar\'{e} sphere}

We give a brief demonstration of the general properties about the
Poincar\'{e} sphere enunciated in Sec. \ref{poincaresphere}. Given
the Poincar\'{e} sphere defined from the uncorrelated basis
($u,v$), we calculate the entanglement and squeezing in different
basis. We recall that the entanglement $\iab$ between modes $a$
and $b$ only depends on the correlation term $|\langle
\daa\dab\rangle|$, whereas the noise sum $\Sigma_{a,b}$ depends on
the quantity $|\langle \daa^2\rangle|^2+|\langle
\dab^2\rangle|^2$. For a given polarization basis
(\ref{baseqcl1}-\ref{baseqcl2}), one has

\beqr |\langle\daa^2\rangle|^2\;& =
&\;\beta^4\langle\dau^2\rangle^2+\alpha^4\langle\dav^2\rangle^2+
2\alpha^2\beta^2\langle\dau^2\rangle\langle\dav^2\rangle\cos2\phi\label{squeez-vs-base1}\\
|\langle\dab^2\rangle|^2\;& =
&\;\alpha^4\langle\dau^2\rangle^2+\beta^4\langle\dav^2\rangle^2+
2\alpha^2\beta^2\langle\dau^2\rangle\langle\dav^2\rangle\cos2\phi
\label{squeez-vs-base2} \eeqr

For $\alpha$ and $\beta$ fixed, $|\langle\daa^2\rangle|$ and
$|\langle\dab^2\rangle|$ are maximal pour $\phi=0\;[\pi]$

\beqr \nonumber|\langle\daa^2\rangle|_{max} =
\beta^2\langle\dau^2\rangle+\alpha^2\langle\dav^2\rangle
\;\;\;and\;\;\;|\langle\dab^2\rangle|_{max} =
\alpha^2\langle\dau^2\rangle+\beta^2\langle\dav^2\rangle\\
\nonumber\Rightarrow\hspace{0.3cm}(|\langle
\daa^2\rangle|+|\langle
\dab^2\rangle|)_{\phi=0}=\max_{a,b}\left[|\langle
\daa^2\rangle|+|\langle \dab^2\rangle|\right]=\langle
\dau^2\rangle+\langle \dav^2\rangle \eeqr

The noise sum is thus minimal for all the linearly polarized modes
with respect to $u,v$, i.e. in the plane ($S'_1,S'_2$)

\beqr\nonumber
\Sigma_{S'_1,S'_2}=\min_{a,b}\Sigma_{a,b}=\langle\delta
X_u^2\rangle_{min}+\langle\delta
X_v^2\rangle_{min}\;\equiv\;\Sigma_{min}\eeqr

If the ellipticity increases ($\phi\neq0$), the noise sum
increases, as can be seen from
(\ref{squeez-vs-base1}-\ref{squeez-vs-base2}). In the meridional
plane ($S'_2,S'_3$) the $a,b$ modes satisfy
($\alpha=\beta=1/\sqrt{2}$)

\beqr
\nonumber(|\langle\daa^2\rangle|^2)_{S'_2,S'_3}=(|\langle\dab^2\rangle|^2)_{S'_2,S'_3}
= \frac{1}{4}\{ \langle\dau^2\rangle^2 + \langle\dav^2\rangle^2 +
2 \langle\dau^2\rangle \langle\dav^2\rangle \cos(2\Phi) \} \eeqr

This quantity is minimal for the circularly polarized modes
$a^*,b^*$ ($\phi=\pi/2$):

\beqr\nonumber|\langle\delta A_{a^*}^2\rangle| + |\langle\delta
A_{b^*}^2\rangle| = |\langle\dau^2\rangle - \langle\dav^2\rangle|
= \min_{a,b}\{|\langle\daa^2\rangle|
+|\langle\dab^2\rangle|\}\label{oscillmin} \eeqr

The noise sum then equals its maximal value $\Sigma_{max}$.
Assuming $\langle\dau^2\rangle\geq\langle\dav^2\rangle$, this
value reads

\beqr\nonumber \Sigma_{a^*,b^*}=\langle\delta
X_{a^*}^2\rangle_{min}+\langle\delta
X_{b^*}^2\rangle_{min}=\langle\delta
X_u^2\rangle_{min}+\langle\delta
X_v^2\rangle_{max}\equiv\Sigma_{max}\eeqr

More generally, one has

\beqr\nonumber \Sigma_{max}=\min \{\;\langle\delta
X_u^2\rangle_{min}+\langle\delta
X_v^2\rangle_{max}\;,\;\langle\delta
X_u^2\rangle_{max}+\langle\delta X_v^2\rangle_{min}\; \} \eeqr

Let us now consider the entanglement. Optimal entanglement is
obtained by construction for modes $a^*,b^*$

\beqr
\nonumber\mathcal{I}_{a^*,b^*}=\min_{a,b}\iab\;=\;\langle\delta
X_u^2\rangle_{min}+\langle\delta
X_v^2\rangle_{min}=\Sigma_{min}\eeqr

The entanglement decreases with the ellipticity. For the linearly
polarized modes ($\phi=0$), Eq. (\ref{correl-ab}) yields

\beqr\nonumber |\langle\daa\dab\rangle|\;=\;\alpha\beta
|\langle\dau^2\rangle-\langle\dav^2\rangle|\eeqr

$\iab$ reaches its maximal value for the $u,v$ modes

\beqr\nonumber \mathcal{I}_{u,v}=\frac{1}{2}\{\langle\delta
X_u^2\rangle_{min}+\langle\delta X_u^2\rangle_{max} +\langle\delta
X_v^2\rangle_{min}+\langle\delta
X_v^2\rangle_{max}\}\equiv\max_{a,b}\iab \eeqr

However, $\iab$ is not constant in the equatorial plane; it is
minimal ($\mathcal{I}_{c,d}=\Sigma_{max}$) for the
$45^{\circ}$-polarized modes ($\alpha=\beta=1/\sqrt{2}$), denoted
by $c,d$.\\

Last, we would like to point out that optimizing the squeezing sum
of two modes is not equivalent to optimizing the squeezing for one
mode only. Finding the maximally squeezed mode is not trivial,
since the $u,v$ modes are not \textit{a priori} independent. No
condition holds on $\langle\dau\dav^{\dagger}\rangle$. The noise
of one quadrature of mode $a$ is given by

\beqr\nonumber \langle\delta X_a^2(\theta)\rangle = \beta^2
\langle\delta X_u^2(\theta)\rangle + \alpha^2 \langle\delta
X_v^2(\theta-\phi)\rangle - 2\alpha\beta\langle \delta X_u(\theta)
\delta X_v(\theta-\phi) \rangle\eeqr

The correlation term can be written as

\beqr\nonumber \langle \delta X_u(\theta) \delta X_v(\theta-\phi)
\rangle&=& \langle \dau\dav^{\dagger}\rangle e^{-i\phi}+ \langle
\dau^{\dagger}\dav\rangle e^{+i\phi}\\
\nonumber&=&2\cos(\phi_C-\phi)|\langle
\dau\dav^{\dagger}\rangle|\eeqr

with $\phi_C$ the phase of $\langle \dau\dav^{\dagger}\rangle $.
The complex general solution takes a simple form if we assume that
$\langle \dau\dav^{\dagger}\rangle $ is a real positive number
($\phi_C=0$). The optimal value is then reached for $\phi=0$ and
one finally gets

\beqr\nonumber
 \min_a \{\langle\delta X_a^2\rangle_{min}\} &=&\frac{1}{2}\{\langle\delta
X_u^2\rangle_{min}+ \langle\delta
X_v^2\rangle_{min}\\
\nonumber &&-\sqrt{(\langle\delta X_u^2\rangle_{min}-
\langle\delta X_v^2\rangle_{min})^2+16\langle
\dau\dav^{\dagger}\rangle^2} \}\eeqr

It is therefore possible to obtain a better squeezing on one mode
than that of $u,v$. Note that, if $\langle
\dau\dav^{\dagger}\rangle=0$, the $u,v$ modes are independent and,
as in our experiments, the best squeezing is that of one of the
$u,v$ modes.

\end{document}